\documentclass[10pt]{iopart}
\usepackage{iopams, graphicx,url, bm, color, tikz, pgfplots, hyperref, soul}
\usepackage[switch]{lineno}
\usepackage[square,sort&compress,numbers]{natbib}
\expandafter\let\csname equation*\endcsname\relax
\expandafter\let\csname endequation*\endcsname\relax
\usepackage{amsmath}
\usetikzlibrary{positioning, shapes, calc}

\let\originalleft\left
\let\originalright\right
\renewcommand{\left}{\mathopen{}\mathclose\bgroup\originalleft}
\renewcommand{\right}{\aftergroup\egroup\originalright}

\hypersetup{colorlinks=true,linkcolor=.,citecolor=blue,urlcolor=blue}

\newcommand{\bmv}{\ensuremath{\bm{v}}}
\newcommand{\bmx}{\ensuremath{\bm{x}}}
\newcommand{\kV}{\ensuremath{\textrm{kV}}}
\newcommand{\m}{\ensuremath{\textrm{m}}}
\newcommand{\cm}{\ensuremath{\textrm{cm}}}
\newcommand{\mm}{\ensuremath{\textrm{mm}}}
\newcommand{\um}{\ensuremath{\mu\textrm{m}}}

\newcommand{\s}{\ensuremath{\textrm{s}}}
\newcommand{\R}{\ensuremath{\textrm{R}}}
\newcommand{\Otwo}{\ensuremath{\textrm{O}_2}}
\newcommand{\Ntwo}{\ensuremath{\textrm{N}_2}}

\newcommand{\OtwoP}{\ensuremath{\textrm{O}_2^+}}
\newcommand{\OtwoM}{\ensuremath{\textrm{O}_2^-}}
\newcommand{\NtwoP}{\ensuremath{\textrm{N}_2^+}}
\newcommand{\NfourP}{\ensuremath{\textrm{N}_4^+}}
\newcommand{\OfourP}{\ensuremath{\textrm{O}_4^+}}
\newcommand{\OtwoPNtwo}{\ensuremath{\textrm{O}_2^+\textrm{N}_2}}
\renewcommand{\e}{\ensuremath{\textrm{e}}^-}
\newcommand{\M}{\ensuremath{\textrm{M}}}
\renewcommand{\O}{\ensuremath{\textrm{O}}}

\begin{document}

\title{3D fluid modeling of positive streamer discharges in air with stochastic photoionization}
\author{Robert Marskar}
\address{SINTEF Energy Research, Sem S\ae lands vei 11, 7034 Trondheim, Norway.}
\ead{robert.marskar@sintef.no}
\date{\today}

\begin{abstract}
  Streamer discharges are thin plasma channels that precede lightning and sparks. They usually evolve in bundles as stochastic tree-like structures and are inherently difficult to model due to their multiscale nature.
  In this paper, we perform a computer investigation of positive streamer discharges in air using contemporary photoionization models.
  We report on three-dimensional computer simulations under conditions that are available in laboratory spark-gap experiments.
  The solutions demonstrate a multiscale morphology consisting of streamer fluctuations, branching, and the formation of a discharge tree.
  Some branches are comparatively thick and noisy, while others are thin, smooth, and carry electric fields exceeding $250\,\textrm{kV/cm}$ at their tips.
  Our results are consistent with past experiments and clarify the puzzling branching dynamics and stochastic morphology of positive streamer discharges. 
\end{abstract}

\maketitle

\ioptwocol

%\linenumbers

\section{Introduction} 
Streamer discharges are fast, filamentary plasma discharges.
They evolve from electron avalanches, are driven by local field enhancement at their tips, and can propagate with or against the electric field.
Streamers that propagate with the electric field are called positive streamers and the ones that propagate against the electric field are called negative streamers.
Streamers are the natural precursor of lightning and sparks, are used in a variety of industrial applications \cite{55956,18870,doi:10.1002/ente.201500127,Boeuf2005,Soloviev2009,Soloviev2017,VanLaer2015, Zhang2018,Wang2018,Dubinova2015, doi:10.1002/2014JD022724, Shi2016, Babich2016, doi:10.1029/2018JD028407}, and also exist naturally as sprite discharges in the upper atmosphere \cite{Fullekrug2006}.
They usually do not appear in isolation but evolve inherently as three-dimensional structures that often branch or form irregular tree structures \cite{VanVeldhuizen2002,Briels2004,Briels2008,Nijdam2010}.
The morphology of a streamer discharge is rich, and their structure is often reminiscent of the fractal-like structure of lightning, but the two are not the same nor do they share the same spatial and temporal scales.
A review of streamer discharges can be found in \citet{Ebert2006}. 

The three-dimensional multiscale structure of streamer discharges restricts analytical treatment and also makes them complicated to model.
Streamer discharges are important in many aspects of plasma physics, but advanced computer modeling is required in order to probe even elementary features.
Streamer modeling mostly fits in two categories: Kinetic and fluid.
Kinetic models evolve electrons as computational (super-)particles and contain all the ingredients for describing streamer discharges.
Various papers report on particle modeling of streamer discharges, but all of these are limited to fairly short propagation lengths \cite{Chanrion2008, Teunissen2016, Fierro2018, Stephens2018, doi:10.1063/1.5019478}.
Fluid models, on the other hand, evolve truncated moments of the electron distribution function.
The computational cost of fluid models is lower than with kinetic models but they are also less accurate, particularly in regions of low electron density.
Fluid models for streamers are widely studied, see e.g. \cite{Hallac2003, Pancheshnyi2008, Luque2008, Papageorgiou2011, Kolobov2012, Nijdam2016a, Plewa2018, Teunissen2017, Marskar2019, Marskar2019a}, but attempts to directly simulate e.g. spark-gap experiments in 3D have so far been unsuccessful due to the numerical effort involved. 

Although widely studied in two spatial dimensions, streamer discharges are inherently three-dimensional and are far from being completely understood.
There is, for example, a current discussion regarding the production of X-rays and terrestrial gamma-ray flashes (TGFs) in laboratory spark experiments and lightning discharges \cite{Dwyer2004, Dwyer2005, Dwyer2005a, Cooray2009, Chanrion2010, Dwyer2012}.
Recently, it has been shown that the TGFs precede the lightning stroke, i.e. TGFs are associated with the lightning leader or streamer corona, and not the stroke \cite{Neubert2019}.
Various branching mechanisms based on hydrodynamic instabilities \cite{Arrayas2002, Ebert2002}, gas impurities \cite{Babaeva2006, Babaeva2008, Babaeva2009a}, and stochastic fluctuations \cite{Teunissen2017, Marskar2019a, Luque2011, Xiong2014, Bagheri_2019} have also been proposed and investigated by the gas discharge community. 

Positive streamers (the type that propagates from the anode to the cathode) also require a source of electrons in front of them.
In air, which we study in this paper, the free electrons are supplied by photoionization of molecular oxygen.
For streamer discharges, photoionization can be modeled with continuum models \cite{Segur2006, Luque2007, Bourdon2007, Janalizadeh2019} or discrete models \cite{Chanrion2008, Xiong2014, Fierro2018, Bagheri_2019}.
Beyond selecting discrete or continuum photons, it is not clear which photoionization model one should use.
In air, the classical Zheleznyak model \cite{1982TepVT..20..423Z, Pancheshnyi2015} has been highly enabling for the streamer discharge community, but it is derived under stationary conditions which are not met for streamer discharges.
More recently, \citet{Stephens2018} has compiled a minimum model for photoionization in air.
We remark that in fluid models the Zheleznyak model is often computed using a Helmholtz-based approach \cite{Segur2006, Luque2007, Bourdon2007, Janalizadeh2019}.
However, if the number of ionizing photons is not sufficiently high, continuum models lead to artificial smoothing of the photoionization profile ahead of the streamer.
Recently, \citet{Bagheri_2019} compared the Zheleznyak model using discrete photons and continuum models.
Although good agreement between the two approaches was found, the results showed a strong sensitivity on the selection of photoionization parameters.
The authors considered a comparatively high and uniform background field and did not include a field-dependent expression for the emission efficiency \cite{Pancheshnyi2015}. 

Motivated by the broad applicability of streamer discharges and their inherent multiscale difficulties, we present a three-dimensional computer investigation of positive streamer discharges in atmospheric air using state-of-the-art photoionization models \cite{Stephens2018}.
Our goal is to shed light on the fundamental multiscale structure of streamer discharges when they are modeled by a fluid model, i.e. how they emerge, propagate, and branch.
In this paper we use a fluid model for all charged species and include ion chemistry in the channel \cite{Pancheshnyi2003}, which is important when modeling timescales longer than some tens of nanoseconds.
We also include supplemental results using the Zheleznyak photoionization model, and remark on some modeling uncertainties relating to the use of both photoionization models.
Our numerical results are computed for $15\,\mm$ pin-plane gaps with an applied voltage of $15\,\kV$, i.e. for an experimentally available configuration.
The results demonstrate streamer morphologies that are consistent with experiments, and thus we are able to clarify the role of photoionization on the evolution and branching of positive streamer discharges in air. 

\section{Physical model}
The fluid streamer model is described by a set of drift-diffusion-reaction equations coupled to the Poisson equation for the electric field:

\begin{eqnarray}
  \label{eq:streamer_model}
  \frac{\partial n_{\textrm{e}}}{\partial t} &= -\nabla\cdot\left(\bmv_{\textrm{e}}n_{\textrm{e}} -D_{\textrm{e}}\nabla n_{\textrm{e}}\right) + S_{\textrm{e}},\\
  \frac{\partial n_i}{\partial t} &= S_i,\\
  \nabla\cdot\bm{E} &= \frac{\rho}{\epsilon_0},
\end{eqnarray}
where $n_{\textrm{e}}$ is the electron density, $\bmv_{\textrm{e}}$ is the electron drift velocity, $D_{\textrm{e}}$ is the electron diffusion coefficient, $S_{\textrm{e}}$ is the electron source term, and $n_i$ is the density of ions of type $i$. $\bm{E}$ is the electric field where $\rho$ is the space charge density.
The ions are assumed to be immobile.

We consider air at standard atmospheric conditions $T=300\,\textrm{K}$, $N=2.45\times 10^{25}\,\m^{-3}$ and use the plasma chemistry as given by \citet{Pancheshnyi2003}, which solves for seven species: electrons, $\textrm{N}_2^+$, $\textrm{O}_2^+$, $\textrm{N}_4^+$, $\textrm{O}_4^+$, $\textrm{O}_2^+\textrm{N}_2$, and $\textrm{O}_2^-$.
The reactions are outlined in \tref{tab:air_reactions}.
In this table, the bottom two reactions denote lumped photoionization reactions which are discussed in detail in \sref{sec:photoionization}.
We also adopt the local field approximation and compute kinetic coefficients by using BOLSIG+\cite{Hagelaar2005,Hagelaar2005a, Phelps1985}.
The computed electron impact and attachment rates are shown in \fref{fig:excitation} as function of the electric field magnitude $E = \left|\bm{E}\right|$.

\begin{table*}[h!t!b!]
  \centering
  \caption{\label{tab:air_reactions} Air reaction mechanism.
    Two-body reactions have units of $\m^3\s^{-1}$ and three-body reactions have units of $\m^6\s^{-1}$.
    Temperatures are in Kelvin.
    The notation $\M$ denotes $\Otwo$ and $\Ntwo$.
    The electron impact rate coefficients and the electron temperature $T_{\textrm{e}} = T_{\textrm{e}}\left(E/N\right)$ are obtained by use of BOLSIG+\cite{Hagelaar2005a} using the PHELPS database \cite{doi:10.1063/1.436700} and the SIGLO databases \cite{Phelps1985}.
    Remaining reaction rates are taken from \citet{Kossyi1992}. }
  \begin{indented}
  \item[]
    \centering
    \begin{tabular}{@{}llll}
      \br
      & Reaction & Rate & Ref. \\
      \mr
      $\R_1$ & $\e + \Ntwo \rightarrow \NtwoP + \e + \e$ & BOLSIG+ & \cite{Hagelaar2005a, Phelps1985} \\
      $\R_2$ & $\e + \Otwo \rightarrow \OtwoP + \e + \e$ & BOLSIG+ & \cite{Hagelaar2005a, doi:10.1063/1.436700} \\
      $\R_3$ & $\NtwoP + \Ntwo + \M \rightarrow \NfourP + \M$ & $5\times 10^{-41}$ &  \cite{Kossyi1992} \\
      $\R_4$ & $\NfourP + \Otwo \rightarrow \OtwoP + \Ntwo + \Ntwo$ & $2.5\times 10^{-16}$ &  \cite{Kossyi1992} \\
      $\R_5$ & $\NtwoP + \Otwo \rightarrow \OtwoP + \Ntwo$ & $1.05\times 10^{-15}T^{-1/2}$ &  \cite{Kossyi1992} \\
      $\R_6$ & $\OtwoP + \Ntwo + \Ntwo \rightarrow \OtwoPNtwo + \Ntwo$ & $8.1\times 10^{-38}T^{-2}$ &  \cite{Kossyi1992} \\
      $\R_7$ & $\OtwoPNtwo + \Ntwo \rightarrow \OtwoP + \Ntwo + \Ntwo$ & $14.8T^{-5/3}\textrm{e}^{\left(-2357/T\right)}$ &  \cite{Kossyi1992} \\
      $\R_8$ & $\OtwoPNtwo + \Otwo \rightarrow \OfourP + \Ntwo$ & $1\times 10^{-15}$ &  \cite{Kossyi1992} \\
      $\R_9$ & $\OtwoP + \Otwo + \M \rightarrow \OfourP + \M$ & $2.03\times 10^{-34}T^{-3.2}$ &  \cite{Kossyi1992} \\
      $\R_{10}$ & $\e + \OfourP \rightarrow \Otwo + \Otwo$ & $2.42\times 10^{-11}\times T_{\textrm{e}}^{-1/2}$ &  \cite{Phelps1985, doi:10.1063/1.436700, Kossyi1992} \\
      $\R_{11}$ & $\e + \OtwoP \rightarrow \Otwo$ & $6\times 10^{-11}T_{\textrm{e}}^{-1}$ &  \cite{Phelps1985, doi:10.1063/1.436700, Kossyi1992} \\
      $\R_{12}$ & $\e + \Otwo + \Otwo \rightarrow \OtwoM + \Otwo$ & $6\times 10^{-39}T_{\textrm{e}}^{-1}$ &  \cite{Phelps1985, doi:10.1063/1.436700, Kossyi1992} \\
      $\R_{13}$ & $\OtwoM + \OfourP \rightarrow \Otwo + \Otwo + \Otwo$ & $1\times10^{-13}$ &  \cite{Kossyi1992} \\
      $\R_{14}$ & $\OtwoM + \OfourP + \M \rightarrow \Otwo + \Otwo + \Otwo + \M$ & $3.12\times 10^{-31}T^{-2.5}$ &  \cite{Kossyi1992} \\
      $\R_{15}$ & $\OtwoM + \OtwoP + \M \rightarrow \Otwo + \Otwo + \M$ & $3.12\times 10^{-31}T^{-2.5}$ &  \cite{Kossyi1992} \\
      $\R_{16}$ & $\e + \Ntwo \rightarrow \e + \Ntwo + \gamma$ & See main text. & See main text. \\
      $\R_{17}$ & $\gamma + \Otwo \rightarrow \e + \OtwoP$ & See main text & See main text. \\
      \br
    \end{tabular}
  \end{indented}
\end{table*}

We incorporate the local electron energy correction due to \cite{Soloviev2014} which accounts for diffusive cooling of electrons near positive streamer tips.
This correction extends the validity of the local field approximation by compensating for a reduction in the average electron energy near strong electron gradients.
The alternative of using an equation for the electron energy is less attractive since it doubles the numerical cost of charged species transport, and also leads to even more stringent constraints on the numerical time steps that can be used. 

Generally, the plasma chemistry of air is quite complicated, and the reaction set that is used here is a minimum reaction set for air \cite{Pancheshnyi2003}.
While there is an abundance of literature on the plasma chemistry of air, we remark that ion conversion $\OtwoP\rightarrow \OfourP$ together with electron-ion recombination $\e+\OfourP \rightarrow \Otwo+\Otwo$ is one of the main electron loss mechanism in the channel, and occurs on the timescale of some tens of nanoseconds.
Some electron loss processes, like dissociative electron attachment, are not included in the present study.
The overall role of incomplete reaction sets for streamer evolution in air is not known. 
We nonetheless remark that \citet{Pancheshnyi2003} have investigated the role of the electron loss processes summarized in \tref{tab:air_reactions}.
The authors found that although the conditions in the channel are influenced by the inclusion or absence of electron loss processes, the streamer discharge dynamics changed insignificantly.

\subsection{Photoionization}
\label{sec:photoionization}
In air, ionizing photons are generated through electron impact excitation of molecular nitrogen to energy levels above the ionization threshold of $\textrm{O}_2$, followed by spontaneous emission.
Photoionization of $\Otwo$ provides a source of free electron-ion pairs around the streamer head. In this model, excited states of nitrogen are first generated through collisions between electrons and molecular nitrogen:

\begin{equation}
  \e + \Ntwo \rightarrow \e + \Ntwo^\ast,
\end{equation}
where $\Ntwo^\ast$ indicates an excited state of molecular nitrogen. The excited nitrogen molecules then terminate through one of three processes: Radiation, predissociation, or collisional quenching, i.e.

\begin{eqnarray}
  &\Ntwo^\ast \rightarrow \varnothing + \gamma, \\
  &\Ntwo^\ast \rightarrow \varnothing, \\
  &\Ntwo^\ast + \M \rightarrow \varnothing,
\end{eqnarray}
where $\gamma$ indicates an ionizing photon, $\M$ denotes either $\Ntwo$ or $\Otwo$, and the target states $\varnothing$ are not of interest and are therefore not tracked further.
Note that predissociation is a fundamental process of the excited state, whereas quenching occurs due collisions with neutral molecules and is therefore a function of gas pressure and temperature.

When ionizing photons are generated, they lead to production of electron-ion pairs through photoionization of molecular oxygen:

\begin{equation}
  \gamma + \Otwo \rightarrow \OtwoP + \e.
\end{equation}

In this paper, radiative transport is treated kinetically since the number of ionizing photons is too low to justify the use of continuum radiative transport models \cite{Segur2006,Luque2007,Bourdon2007,Janalizadeh2019}.
As mentioned above, we consider two photoionization models, the minimal model suggested by \citet{Stephens2018} and the classical Zheleznyak model \cite{1982TepVT..20..423Z}. 

\subsubsection{\citet{Stephens2018} photoionization model}
\label{sec:stephens}
\begin{figure}[h!t!b!]
  \centering
  \includegraphics{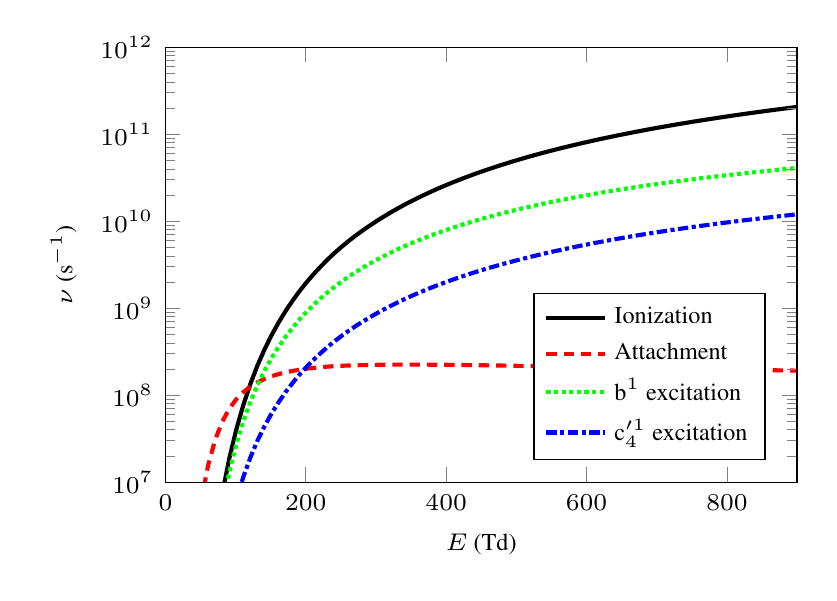}
  \caption{Electron impact ionization, attachment, and excitation frequencies that are used in the \citet{Stephens2018} model.}
  \label{fig:excitation}
\end{figure}

We first outline the \citet{Stephens2018} photoionization model which includes eight radiative transitions from three excited states of molecular nitrogen.
The electronically excited states ($\Ntwo^\ast$) are the singlet states $\textrm{N}_2\left(\textrm{c}_4^{\prime 1}\Sigma_u^+\right)$, i.e. the Carrol-Yoshino band, and the Birge-Hopfield II band $\textrm{N}_2\left(\textrm{b}^{1}\Pi_u^+\right)$.
Evaluations of predissociation lifetimes, radiative lifetimes, and excitation probabilities \cite{Stephens2018} indicate that the vibrational levels $c_4^{\prime 1}\Sigma_u^+\left(\nu^\prime = 0,1\right)$ and $b^{1}\sum_u^+\left(\nu^\prime=1\right)$ produce the majority of ionizing photons.
Numerically, the number of ionizing photons of type $\gamma$ emitted by each excited state in a computational discretization volume $\Delta V$ and time step $\Delta t$ is determined as follows:
\begin{enumerate}
\item Determine the number $N_r$ of radiative de-excitations by drawing from a Poisson distribution with mean $\nu n_{\textrm{e}}\zeta p_r\Delta V\Delta t$, where $\nu$ is the excitation rate, $\zeta$ is the relative excitation probability, and $p_r = k_r/(k_q+k_p+k_r)$ where $k_r$ is the total radiation rate of the excited state, $k_p$ is the predissociation rate, and $k_q$ is the collisional quenching rate.
  The excitation rates $\nu$ are field-dependent and are computed using data from \citet{Malone2012} and are shown in \fref{fig:excitation}. 
\item Determine the number $N_{r, \gamma}$ of each photon type emitted by drawing $N_r$ trials from a multinomial distribution with probabilities $p_\gamma = k_\gamma/(\sum_{\gamma}k_{\gamma})$ where the sum runs over the radiative transitions associated with the excited state.
  Then determine the final number of ionizing photons by drawing $N_{r, \gamma}$ trials from a binomial distribution with success probability (i.e. photoionization efficiency) $\xi_\gamma$.
  Only these photons are tracked kinetically\footnote{
    The process can be simplified by assuming that each photo-generation process is independent, in which case the photons can be sampled directly from Poisson distributions with parameters $\nu\zeta\xi_\gamma n_e k_{r,\gamma}/(k_q + k_p + k_r)\Delta V\Delta t $.
}.
\end{enumerate}

The photoionization parameters required for this process have been summarized by \citet{Stephens2018} in terms of cross sections for general $\Otwo$-$\Ntwo$ mixtures.
These cross-sections have been used to compute the photoionization parameters for atmospheric air as discussed above, and they are presented in \tref{tab:photoionization}.

The quenching rates for the singlet states of $\textrm{N}_2$ are, unfortunately, not known with desired accuracy.
Pending an improvement in the measurement of quenching rates for $\textrm{c}_4^{\prime 1}\Sigma_u^+$ and $\textrm{b}^1\Pi_u$, we have taken an approximate value $k_q/N = 10^{-16}\,\textrm{m}^3/\textrm{s}$, or $k_q = 2.45\times 10^{9}\,\textrm{s}^{-1}$.
We have taken a slightly lower value than suggested by \citet{Pancheshnyi2015} in order to compensate for omission of other photoionization mechanisms.
Note that the collisional quenching rate exceeds the predissociation rates by about one order of magnitude for the $\textrm{c}_4^{\prime 1}\Sigma_u^+\left(0\right)\rightarrow X^1(0,1)$ and $\textrm{b}^1\Pi_u(1)\rightarrow X^1(0,1)$ transitions. 

\begin{table*}[h!t!b!]
  \centering
  \caption{\label{tab:photoionization} Atmospheric air photoionization mechanism of molecular oxygen through excited states of molecular nitrogen ($\textrm{N}_2$) using the \citet{Stephens2018} model.
    Each row of the table represents one ionizing radiative transition.
    The quenching rate constant for atmospheric pressure air is taken as $k_q = 2.45\times 10^{9}\,\textrm{s}^{-1}$. }
  \begin{indented}
  \item[]
    \centering
    \begin{tabular}{@{}llllll}
      \br
      Transition $\gamma$                                               & $k_p\,(\times 10^8\,\textrm{s}^{-1})$   & $k_{r,\gamma}\,(\times 10^8\,\textrm{s}^{-1})$   & $\kappa^{-1}\,(\um)$  & $\xi$ & $\zeta$ \\
      \mr
      $\textrm{c}_4^{\prime 1}\Sigma_u^+\left(0\right)\rightarrow X^1(0)$ & 2.39    & 11.2         & $19$         & $0.23$  & 0.8\\
      $\textrm{c}_4^{\prime 1}\Sigma_u^+\left(0\right)\rightarrow X^1(1)$ &         & 1.89         & $146$        & $0.77$ & \\
      \mr
      $\textrm{c}_4^{\prime 1}\Sigma_u^+\left(1\right)\rightarrow X^1(0)$ & 17.7    & 0.41         & $30$         & $0.045$ & 0.034\\
      $\textrm{c}_4^{\prime 1}\Sigma_u^+\left(1\right)\rightarrow X^1(1)$ &         & 6.97         & $21$         & $0.10$ & \\
      $\textrm{c}_4^{\prime 1}\Sigma_u^+\left(1\right)\rightarrow X^1(2)$ &         & 4.06         & $58$         & $0.75$ & \\
      $\textrm{c}_4^{\prime 1}\Sigma_u^+\left(1\right)\rightarrow X^1(3)$ &         & 1.45         & $158$        & $0.69$ & \\
      \mr
      $\textrm{b}^1\Pi_u(1)\rightarrow X^1(0)$                         & 4.68    & 0.42         & $236$        & $0.72$ & 0.014\\
      $\textrm{b}^1\Pi_u(1)\rightarrow X^1(1)$                         &         & 0.80         & $298$        & $0.69$ & \\
      \br
    \end{tabular}
  \end{indented}
\end{table*}

The present photoionization model disregards radiation trapping through self-absorption.
We remark that radiation trapping is primarily relevant for transitions where the emitted radiation is resonant with a radiative re-excitation from highly populated state. 
At room temperature only the ground state of molecular is populated so that trapping is most relevant for the $c_4^{\prime 1}\Sigma_u^+\left(\nu^\prime = 0,1\right) \rightarrow X^1(0)$ and $\textrm{b}^1\Pi_u(1)\rightarrow X^1(0)$ transitions.
Although it would be desireable to incorporate radiation trapping, it also requires invasive changes to our numerical approach since the excited species of nitrogen must then be tracked kinetically.
We also ignore the time-of-flight of the photon, an approximation which is made for simplicity rather than necessity. 

\subsubsection{Zheleznyak photoionization model}
\label{sec:zheleznyak_model}
\begin{figure}[ht]
  \centering
  \includegraphics{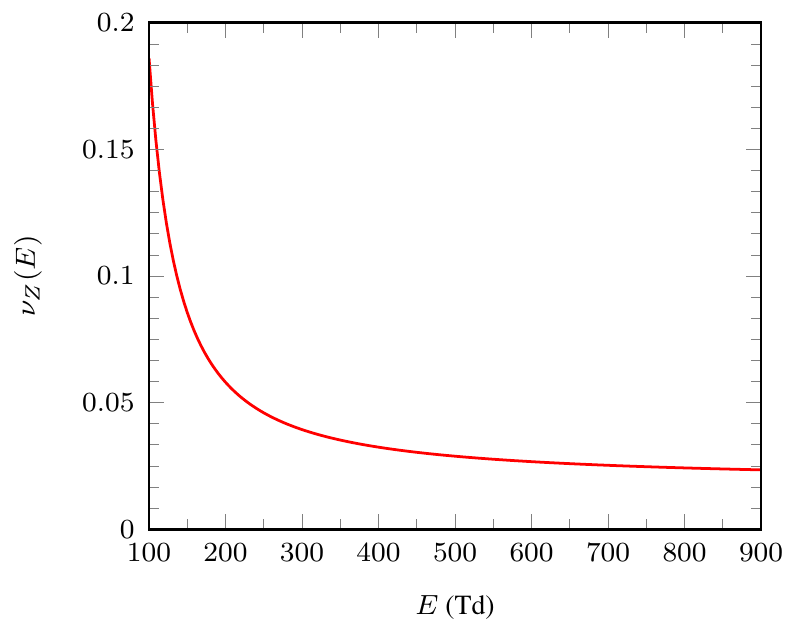}
  \caption{
    Fit factor for the Zheleznyak model for photoionization \citet{Pancheshnyi2015}.
    The curve indicates how many ionizing photons are generated per electron impact ionization event (in the absence of quenching). }
  \label{fig:zheleznyak_function}
\end{figure}

Next, we discuss the Zheleznyak model for photoionization with discrete photons, using the improvements provided by \citet{Pancheshnyi2015}.
In this model the photons are generated at an average volumetric rate of

\begin{equation}
  R_Z = \frac{p_q}{p + p_q}\nu_{Z}\left(E\right)\alpha n_{\textrm{e}}|\bm{v}_{\textrm{e}}|,
\end{equation}
where the quenching pressure $p_q = 40\,\textrm{mbar}$ accounts for collision quenching, $\nu_Z\left(E\right)$ is a lumped function that accounts for both excitation efficiencies and photoionization probabilities, and $\alpha$ is the Townsend ionization coefficient.
\citet{Pancheshnyi2015} has provided a fit of $\nu_Z(E)$ to experiments, which is shown in \fref{fig:zheleznyak_function} for atmospheric air.

In the Zheleznyak photoionization model discrete photons are drawn at every time step from a Poisson distribution with parameter $R_Z\Delta V\Delta t$.
The frequency of each photon is drawn randomly from an interval $f\in[f_1, f_2]$ where 

\begin{eqnarray}
  f_1 &= 2.925\,\textrm{PHz}, \\
  f_2 &= 3.059\,\textrm{PHz}.
\end{eqnarray}
The propagation distance is then drawn from an exponential distribution with parameter

\begin{equation}
  \mu = K_1\left(\frac{K_2}{K_1}\right)^{\frac{f-f_1}{f_2-f_1}},
\end{equation}
where $K_1 = 530\,\m^{-1}$ and $K_2 = 3\times 10^4\,\m^{-1}$.
These values are valid for atmospheric air only.
For scaling to other $\Ntwo$-$\Otwo$ mixtures or different pressures, see e.g. \cite{Liu2004}.

\subsubsection{Comparison between photoionization models}
\begin{figure}[h!t!b!]
  \centering
  \includegraphics{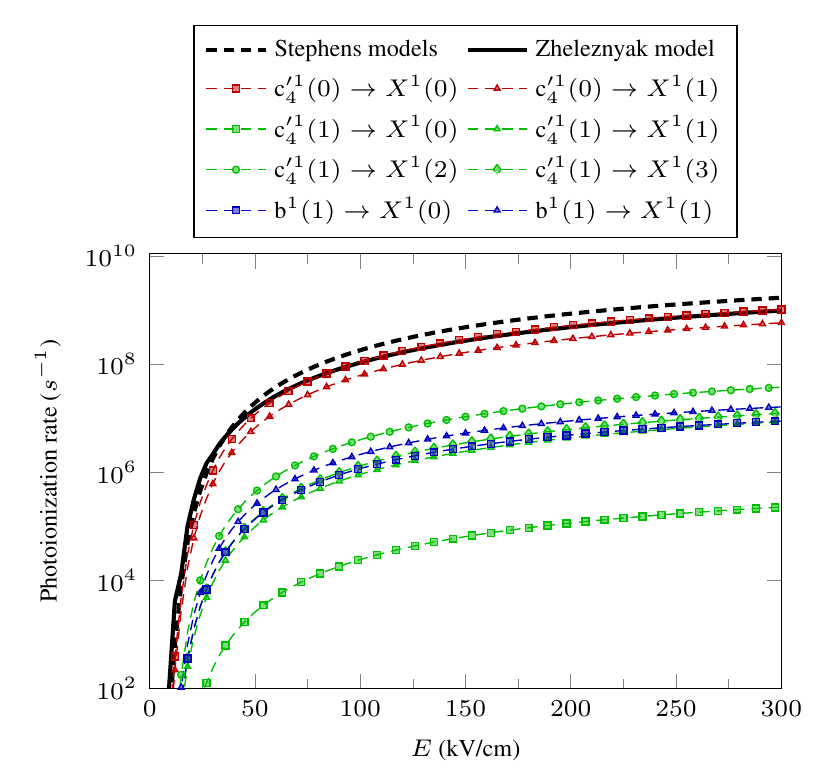}
  \caption{Comparison between the photoelectron production rate in atmospheric air using the \citet{Stephens2018} model and the Zheleznyak model. The broken lines with markers show individual contributions in the \citet{Stephens2018} model.}
  \label{fig:photoelectron_production}
\end{figure}

To obtain a quantitative understanding of the number of photoelectrons that are produced in a discharge, we compare the normalized photoionization rate predictions of the two models.
The curves in \fref{fig:photoelectron_production} show the photoionization rate, i.e. the curve for the Zheleznyak model shows $R_z/n_{\textrm{e}}$ and the curves for the \citet{Stephens2018} model show $\nu\zeta\xi_\gamma k_{r,\gamma}/(k_q + k_p + k_r)$.
We find that the two models agree quite well in terms of the total number of photoelectrons that are produced. 
Here, the model by \citet{Stephens2018} produces approximately a factor of two more photoelectrons than the Zheleznyak model, but we remark that this factor depends on the selection of the quenching rate.

Although we have not undertaken a study that attempts to classify the importance of each radiative transition in a streamer discharge, we observe that a major fraction of photoelectrons appear due to radiative de-excitations of the $c_4^{\prime 1}\Sigma_u^+\left(\nu^\prime = 0\right)$ state.
The average absorption distances of the $c_4^{\prime 1}\Sigma_u^+(0)\rightarrow X^1(0,1)$ transitions are approximately $19\,\um$ and $150\,\um$. 
In the Zheleznyak model the photons are more evenly distributed, with the average absorption distance being approximately $500\,\um$ in atmospheric air.
The Zheleznyak model therefore overestimates the number of long-range photons, and underestimates the number of short-range photons.

\subsection{Numerical discretization} 
\label{sec:discretization}
The above equations are solved with an embedded boundary adaptive mesh refinement (EBAMR) methodology by using Chombo \cite{chombo, ebchombo}.
The EBAMR technology uses a combination of Cartesian Adaptive Mesh Refinement (AMR) and embedded boundaries (i.e. ''cut-cells'') that enable numerical representations of electrodes and other internal boundaries.
Special discretization techniques are used to handle the spatial and temporal discretization of Cartesian cut-cells \cite{Marskar2019a}. 

\begin{figure}[ht]
  \centering
  \includegraphics[width=.4\textwidth]{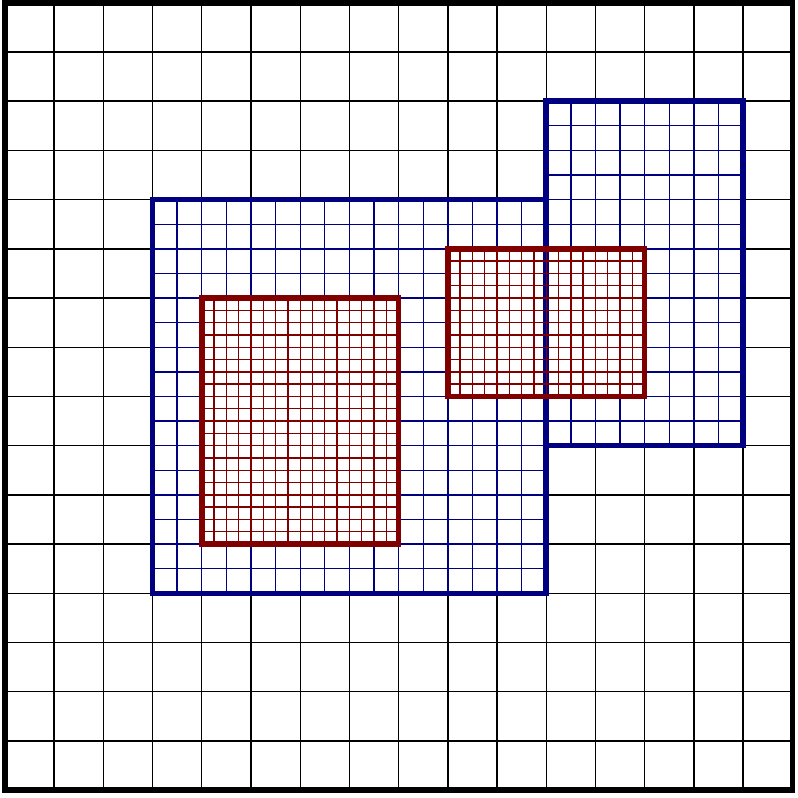}
  \caption{Classic cartoon of AMR grids. The coarsest grid covers a domain of $16\times16$ cells with two-levels of properly nested refined grids.
    Bold lines represent grid boundaries.
    Note that the grids do not have unique parent-children relationships.
    For example, one of the finest-level grids has two parents.}
  \label{fig:amr}
\end{figure}

\Fref{fig:amr} shows a classical cartoon of AMR grids away from solid boundaries. With AMR, grids simultaneously exist at multiple levels and the numerical discretization is matched at refinement boundaries.
The use of AMR is mandatory for large scale 3D simulations since covering the domain with a fine grid leads unrealistic demands on computational hardware.
Instead, AMR allows grids to be created and destroyed during the simulation in such a way that the grid dynamically adapts to the numerical solution as it evolves. 

Following a previous paper \cite{Marskar2019a}, we use a second order spatial discretization using finite volumes.
However, in this paper we use a (formally) first order Godunov splitting for advancing $t_k \rightarrow t_k+\Delta t = t_{k+1}$ as follows:

\begin{eqnarray}
  \label{eq:godunov}
  \psi &= n^{k} - \Delta t\nabla\cdot\left(\bmv_{\textrm{e}}^kn^{k+1/2} - D_{\textrm{e}}^k\nabla n^k\right),\\
  \bm{E}^{k+1} &= \bm{E}\left(\psi\right), \\
  \label{eq:chemistry}
  \psi &\xrightarrow{S, \Delta t} n^{k+1},
\end{eqnarray}
where the divergence operators are now placeholders for more complicated discretization procedures \cite{Marskar2019a}.
The coefficients $\bm{v}_{\textrm{e}}^k$ and $D_{\textrm{e}}^k$ are computing using $\bm{E}^{k}$ and the symbolic notation $\xrightarrow{S, \Delta t}$ indicates a solution to the reactive problem $\partial_tn = S$ over a time step $\Delta t$ using a second order Runge-Kutta method with $\psi$ as the initial solution.
$\bm{E}\left(\psi\right)$ indicates the solution to the Poisson equation with space charge $\rho = \rho\left(\psi\right)$.
This solution is computed using a geometric multigrid method.
Specifically, we use red-black Gauss-Seidel relaxation on each grid level and a biconjugate gradient stabilized solver in the bottom of the V-cycle.
A solver tolerance of $10^{-10}$ on the max-norm for the electric potential is used as an exit criterion for the multigrid solver.

A beneficial side-effect of the splitting in \eref{eq:godunov} is the semi-implicit coupling to the plasma chemistry terms since the chemistry step \eref{eq:chemistry} does not generate space charge and $\bm{E}\left(\psi\right) = \bm{E}\left(n^{k+1}\right)$.
Our empirical experience indicates that splitting the transport and chemistry leads to favorable stability properties, also for cases that are not investigated here (such as propagation of streamers along dielectrics). 
The advective term is partially centered at time $t_k+\Delta t/2$, which is done by extrapolating the face-centered states that are required for the finite-volume discretization to half times.
This reduces the advective discretization error and permits use of higher Courant-Friedrichs-Lewy (CFL) numbers without compromising the solution quality \footnote{Observe that for evolution in the Laplacian field and in the absence of diffusion and plasma chemistry, the discretization is second order accurate.}.
Expressions for the advective slopes are found in \cite{trebotich2015}.
We include the source and diffusion terms in the time-extrapolation.
This leads to a partial upwinding of the source term and enhances the stability of the scheme (see \cite{Villa2014} for a related numerical technique).
The combination of Godunov splitting with state time-extrapolation yields a comparatively stable scheme which is suitable even for simulating near-stagnant streamers \cite{Pancheshnyi2004}, which are numerically challenging.
After updating \eref{eq:godunov} we deposit the photons on the mesh, and update boundary conditions and kinetic coefficients. 

Time steps are limited by standard CFL constraints on advection and diffusion.
We have not made efforts to eliminate the dielectric relaxation time, but remark that this is possible by computing the transport step semi-implicitly \cite{Ventzek1994}.

Ionizing photons are given a random propagation direction on the unit sphere and the absorption distance is randomly drawn from an exponential distribution.
Photoemission is disregarded in this paper, so if a photon trajectory intersects an internal boundary (e.g. an electrode) or a domain boundary, it is removed from the simulation.
Photon generation and transport is instantaneous and the photons are deposited on the mesh with a cloud-in-cell (CIC) scheme.
Generally, photons are deposited on the finest grid level that is available.
An exception to this rule is made on the fine side of refinement boundaries.
In order to prevent deposition clouds that hang into ghost cells over refinement boundaries, photons that live in the first strip of cells on the fine side of a refinement boundary are instead deposited on the coarse grid, and the result is then interpolated to the fine grid before depositing the rest of the fine-grid photons.

When discrete photons are used, photoionization events lead to appearance of patches of fluid densities on the grid.
For the electrons, these patches numerically and physically diffuse on the grid and become smeared out over several grid cells.
The combined model therefore still understimates the spatial noise in electron distribution ahead of the streamer.

\subsection{Simulation conditions}
\label{sec:conditions}
\begin{figure}[h!t!b!]
  \centering
  \includegraphics{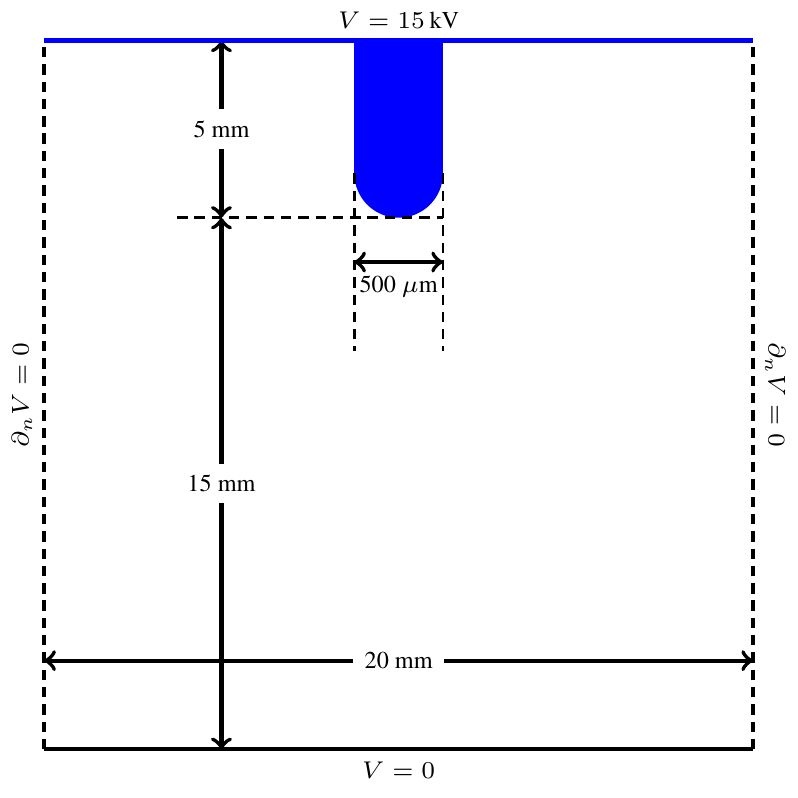}
  \caption{2D cross-sectional view of the simulation domain, also showing the electrostatic boundary conditions. The radius of the pin electrode is not to scale. }
  \label{fig:domain}
\end{figure}

We consider a $(2\,\cm)^3$ domain with a vertical plane-protrusion geometry.
A $5\,\mm$ long cylindrical electrode with a spherical cap at the end sticks out of the live electrode plane, whereas the opposite plane is grounded.
The electrode diameter is $500\,\um$, and the distance between the live electrode and the ground plane is $15\,\mm$.
Homogeneous Neumann boundary conditions are placed on the side edges, and all simulations start from a voltage $V_0=15\,\kV$ on the electrode.
A cross section of the computational domain and the boundary conditions are shown schematically in \fref{fig:domain}. 

The electrode is numerically represented by a level-set function $f(\bm{x})$ where $f(\bm{x}) = 0$ indicates the surface of the electrode.
Positive and negative values of this function indicate points inside and outside of the object, respectively.
Here, this function is created using constructive solid geometry by taking the union of a cylinder and sphere.
At the beginning of the simulation, the geometric quantities of the cut-cells (e.g volume fraction) are found by computing the intersection of the Cartesian grid cells with this function.

For initial conditions, we disregard preionization and use a small quasi-neutral initial plasma patch of electron-ion pairs (80\% $\textrm{N}_2^+$ and 20\% $\textrm{O}_2^+$),

\begin{equation}
    n_{\textrm{e}}=n_0\exp\left(-\frac{\left(\bmx-\bmx_0\right)^2}{2R^2}\right),
\end{equation}
where $R$ is the patch radius and $\bmx_0$ is the electrode tip, $n_0 = 10^{16}\,\m^{-3}$ and $R=100\,\um$.
We have chosen to use smooth and symmetric initial conditions in order to ensure that any stochasticity that develops is not a result of initial perturbations.
Physically correct initial conditions could be provided by for example depositing initial electron particles on the grid as fluid densities. 

The equations are solved over an EBAMR grid hierarchy with a base level of $160^3$ cells and 6 refinement levels with factor two refinement between each level.
The effective grid size is therefore $(10240)^3$ and the finest possible grid resolution in the simulation domain is roughly $2\,\um$.
The results in this paper use a tiled-based mesh generator \cite{Gunney2016} instead of the classical Berger-Rigoutsos algorithm \cite{Berger1991}.
Tiled patch-based grids look similar to tree-based grids but are more general since the domain extensions are not restricted to being an integer power of 2 (e.g. $2^{13}$ cells), and can also use refinement factors other than 2.
The tile size is set to 16, i.e. on each AMR level the mesh represented by a union of grids with $16^3$ cells. 

The mesh is refined if

\begin{equation}
  \alpha\Delta x > 1\textrm{ or } \left|\nabla E\right|\Delta x/E > 5,
\end{equation}
and coarsened if

\begin{equation}
  \alpha\Delta x < 0.2\textrm{ and } \left|\nabla E\right|\Delta x/E < 2,
\end{equation} 
where $\Delta x$ is the grid spacing.
The first refinement criterion $\alpha \Delta x > 1$ is used to trigger refinement where impact ionization occurs, and the second refinement criterion $\left|\nabla E\right|\Delta x/E > 5$ is used in order to ensure that the electric field is resolved with a sufficient resolution.
We always include a buffer zone of $8$ cells when refining the grid.
This means that if any cell is flagged for refinement within a tile, we also refine the neighboring tiles.
In the simulations the CFL number is set to 0.8.

\begin{figure*}[h!t!b!]
  \centering
  \includegraphics{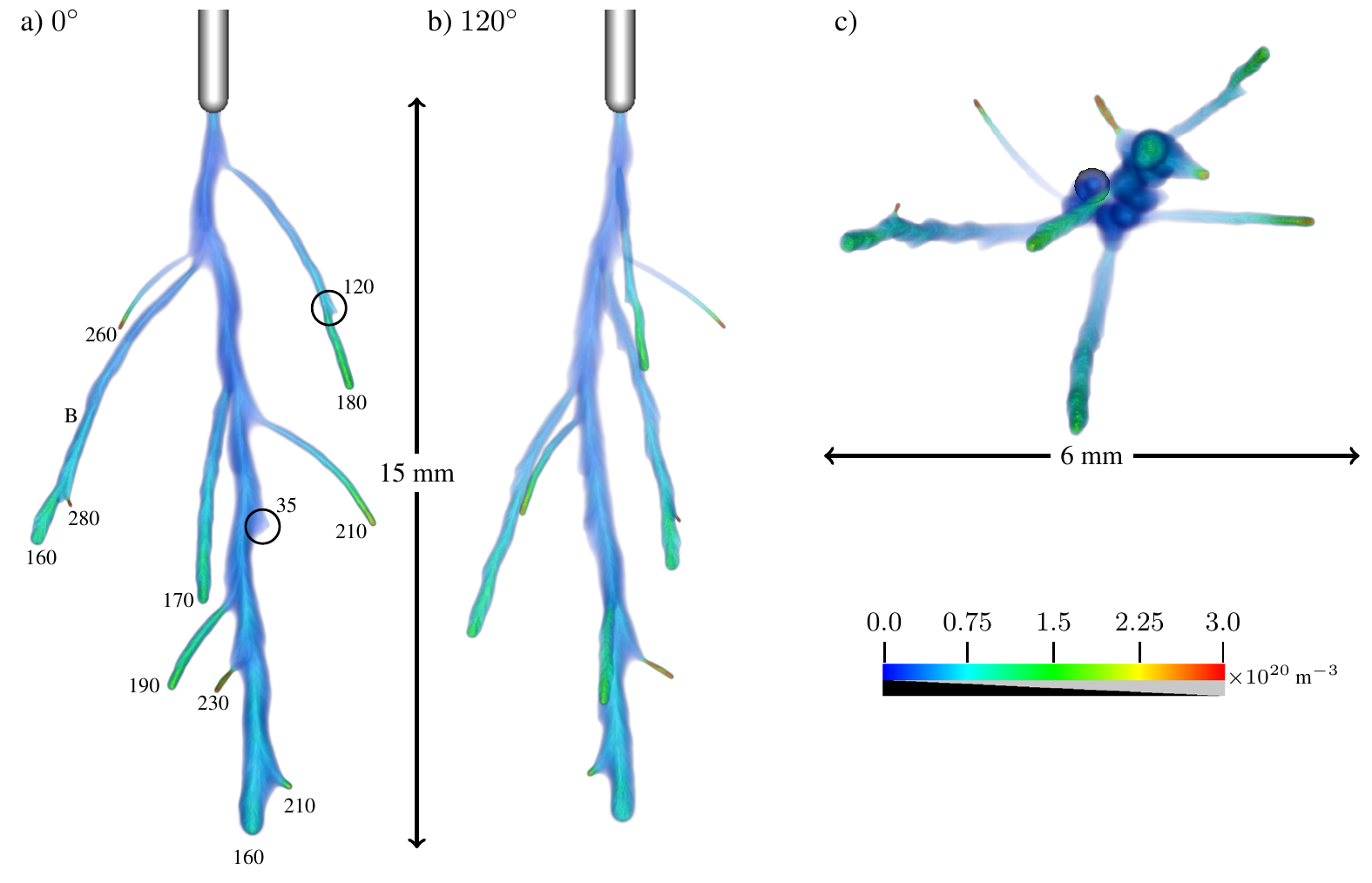}
  \caption{Rotated views of the electron density after $42\,\textrm{ns}$.
    The numbers in figure a) indicate the electric field at the streamer tips in units of kV/cm.
    The solid circles in a) indicate branches that failed to fully develop into streamers, but left behind a high-field region.
    Figure b) is rotated $120^\circ$ degrees with respect to figure a).
    Figure c) views the streamers from below.
  }
  \label{fig:15kv_15mm_42ns}
\end{figure*}

Although three-dimensional computer simulations of streamer discharges are generally speaking quite expensive, the use of AMR leads to drastic reductions in simulation times.
For example, the most time consuming simulation in this paper completed in just less than two days on 512 cores.
The hardware consisted of 16 nodes of dual Intel Xeon E5-2683v4 cores, which by current standards is fairly old hardware.
The simulation took around 28,000 time steps using up to 100 million grid cells for a total simulation time of $42\,\textrm{ns}$, and generated about 60~TB of plot and checkpoint data.
Regridding was performed every ten time steps and accounts for approximately 10\% of the total simulation time.
Parallel I/O with HDF5 was also performed every ten time steps, and accounted for around 20\% for the simulation time.
For our simulations, discrete photons are computationally far cheaper than continuum models since there are comparatively few photons generated at every time step, and there is no need to solve multiple elliptic equations as is done with the Helmholtz approach \cite{Segur2006, Luque2007, Bourdon2007, Janalizadeh2019}.

\section{Results with the \citet{Stephens2018} photoionization model}
\label{sec:results}
First, we present results computed with the \citet{Stephens2018} photoionization model.
These results are presented in \fref{fig:15kv_15mm_42ns} and demonstrate morphological features that are observed in experiments, for example streamer fluctuations and branching.
Animations of this simulation are provided in the supplemental material.
The animations provide a clearer picture of the discharge morphology, as well as better illustrations of the relative filament velocities and branching processes. 

We do not observe inception clouds \cite{Briels2008} in our simulations, which we attribute to our use of a fairly blunt electrode tip.
The use of a sharper electrode tip or other boundary conditions (e.g. grounded side walls) yields higher field strengths near the anode and can lead to appearance of inception clouds. Experimentally \cite{Briels2008,Briels2008a}, inception clouds are seen for sharp tips, high voltages, or low pressures, but none of these conditions are met in our simulations. 

In \fref{fig:15kv_15mm_42ns} we observe 10 streamer branches.
The thinnest streamers are about $100\,\um$ in width whereas the thickest one is around $500\,\um$.
The streamers with the smallest radii propagate with velocities $v \sim 0.15\,\textrm{mm/ns}$ while the thickest streamer propagate with velocity $v \sim 0.33\,\textrm{mm/ns}$.
Streamer radii and velocities agree with experimental observations performed at similar (but not identical) conditions \cite{Briels2008a}. 

The branching distance is significantly different for different branch thicknesses.
For example, the ''main stem'' in \fref{fig:15kv_15mm_42ns} has a branching distance $D$ relative to its diameter $d$ of $D/d\approx 4$, whereas the branch labeled B has in \fref{fig:15kv_15mm_42ns}a) has $D/d\approx 30$.
In experiments \cite{Briels2008a}, thick streamers have $D/d=8\pm4$ while for thin (but not minimal) streamers $D/d=11\pm4$, but there are generally large variations in the experimentally observed branching distance.
For example, streamers don't branch below a certain diameter. 

Thinner streamers have higher fields at their tips.
The analytical model suggested by \citet{Chen2013} estimates a maximum electric field for a streamer with radius $R_s$ as $E_{\textrm{max}}\approx U/(k R_s)$ where $U$ is voltage in the streamer head and $k=\textrm{2-4}$ is a correction factor for air.
Voltage depletion in the channel is about $5\,\textrm{kV/cm}$ in atmospheric air, so that after $1\,\cm$ propagation we obtain the estimate $E_{\textrm{max}}\approx\,\textrm{250-500kV/cm}$ for $R_s=100\,\um$.
In the simulations we observe transient fields up to $300\,\textrm{kV/cm}$ at the branching point and sustained fields at the streamer tips up to $260\,\textrm{kV/cm}$.

\begin{figure}[h!t!b!]
  \centering
  \includegraphics{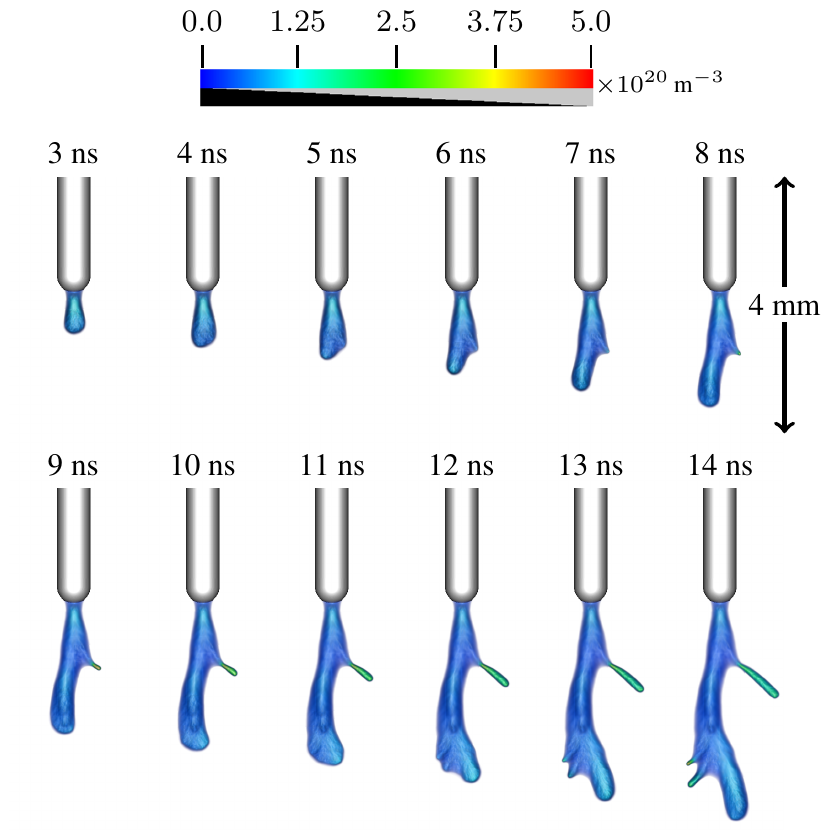}

  \caption{Multiscale branching process and emergence of thin streamers in the main stem wake.
    The figure shows a closeup of the discharge evolution in \fref{fig:15kv_15mm_42ns}.
    Branching occurs due to a deflection of the streamer path that leaves behind a curved region in the space charge layer, particularly visible for times 4ns-8ns, from which a thin streamer develops.}
  \label{fig:branching}
\end{figure}

Next, we turn to streamer branching which is computationally observed to occur as follows: Photoionization around the streamer head yields stochastically distributed electron avalanches that propagate towards the streamer tip.
As the electric field gravitates towards higher plasma densities in front of the streamer, noise or uneven distributions in the incoming avalanches can result in a shift of the electric field maximum slightly away from the streamer tip.
This leads to spatial variations in the electron density inside the streamer, but can also lead to macroscopic variations such as fluctuations in streamer radius or deflection of the original trajectory.
If the reaction zone center is shifted and advances slightly further (but off-axis), the space charge layer curvature and electric field magnitude decreases at the point where the streamer changed trajectory.
This essentially leads to a flattening of the space charge layer and a decrease in the field magnitude where the layer is less curved.
Ultimately, additional field maxima then appear in the streamer head, and these act as nuclei for new branches.
The branching process is shown in \fref{fig:branching}. 

We have not observed branching from filaments that do not fluctuate, further indicating that these two phenomena are linked.
Also note that even when additional local field maxima emerge in the streamer head, there is no guarantee that they all lead to new streamer branches.
When fluctuations are strong but branches fail to grow, a ''kink'' in the space charge layer in the streamer wake appears instead.
The kink, or failed streamer branch, consists of a small cluster of positive ions that sticks out from the space charge layer and outside of it we find a small-volume region with an electric field well above the critical field.
Such kinks have been observed computationally.
They are indicated by black circles in \fref{fig:15kv_15mm_42ns}a), and for both indicated regions the field in the streamer wake is above the ionization level of the gas but no streamers emerge.
Analogous structures probably exist also for negative streamers.
Whether or not the ''failed branches'' emit any observable light after they stop, or if they develop into highly latent branches, is an open question that will be investigated in the future.
If they do emit light, they may be related to streamer beads \cite{Stenbae-Nielsen2000, Cummer2006, Luque2011a, Luque2016, Kochkin2016}. 

\begin{figure}[h!t!b!]
  \centering
  \includegraphics{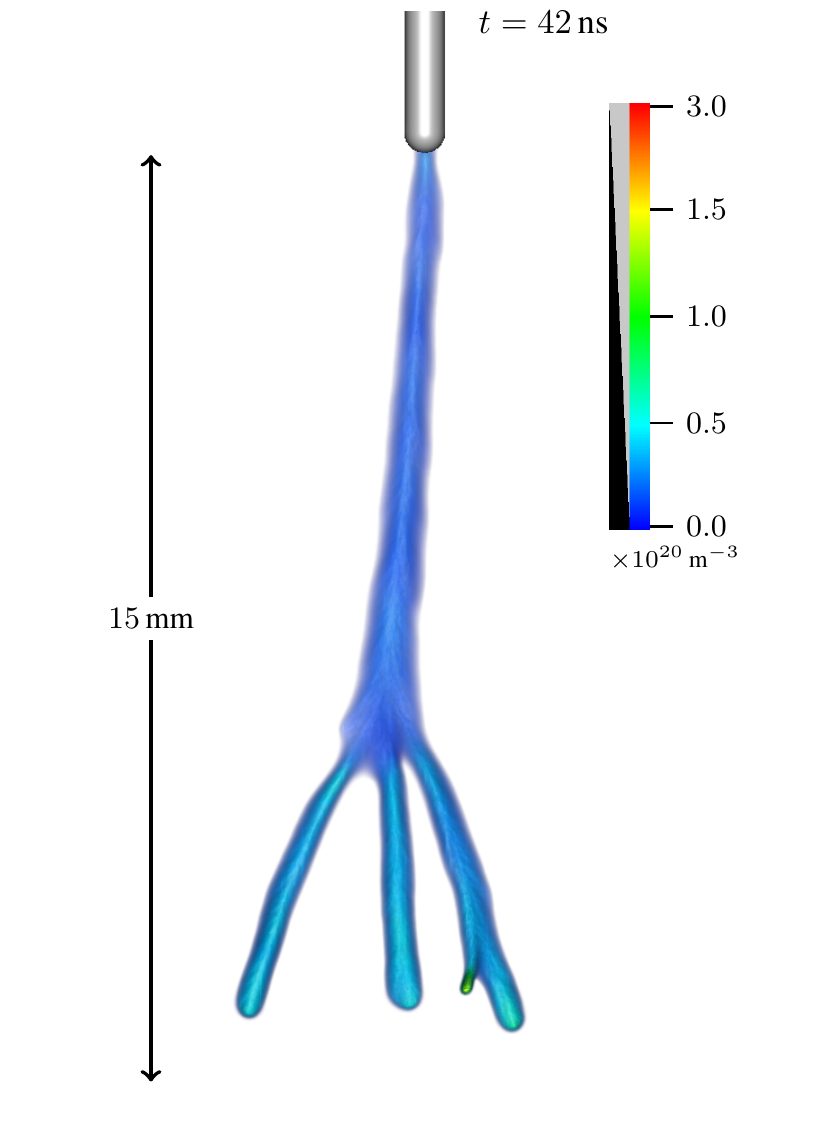}
  \caption{(Color online) Electron densities computed using the \citet{Stephens2018} model with an artificially lowered collisional quenching rate.}
  \label{fig:loQuench}
\end{figure}

\subsection{Results with lower quenching rate constants}
To illustrate the role of collisional quenching, we now provide results computed with an artifically lowered quenching rate.
For the \citet{Stephens2018} photoionization model we have $k_q \gg k_{r,\gamma}$ for the $\textrm{b}^1\Pi_u(1)\rightarrow X(0,1)$ transitions whereas we have $k_q\sim k_{r,\gamma}$ for the $\textrm{c}_4^{\prime 1}\Sigma_u^+\left(0\right)\rightarrow X^1(0)$ transitions.
These inequalities rely on the use of a single quenching rate constant for both singlet states, but the trend is nonetheless clear: Excited states with long radiative lifetimes are quenched more efficiently.
We now set $k_q/N = 10^{-17}\,\textrm{m}^3/\textrm{s}$ which lowers the quenching rate by one order of magnitude such that the predissociation and quenching lifetimes have the same order of magnitude.
Note that we artificially lower $k_q$ but leave the other reaction rates untouched. 
At lower quenching levels less branching is expected since the noise in the ionization level in front of the streamer is reduced due to appearance of more photoelectrons, particularly from the $\textrm{c}_4^{\prime 1}\Sigma_u^+\left(0\right)\rightarrow X^1(0,1)$ and $\textrm{b}^1\Pi_u(1)\rightarrow X^1(0,1)$ transitions.
The relative increase in the number of photons from the $\textrm{c}_4^{\prime 1}\Sigma_u^+\left(1\right)\rightarrow X^1(0,1,2,3)$ is less since the excited state is already heavily predissociated.
\Fref{fig:loQuench} shows the results computed using the \citet{Stephens2018} model for photoionization with the lowered quenching rate.
The results show substantially less branching.
We attribute this result to suppression of stochastic fluctuations in front of the streamer, which occurs due to an increase in the photoionization rate. 

\begin{figure}[h!t!b!]
  \centering
  \includegraphics{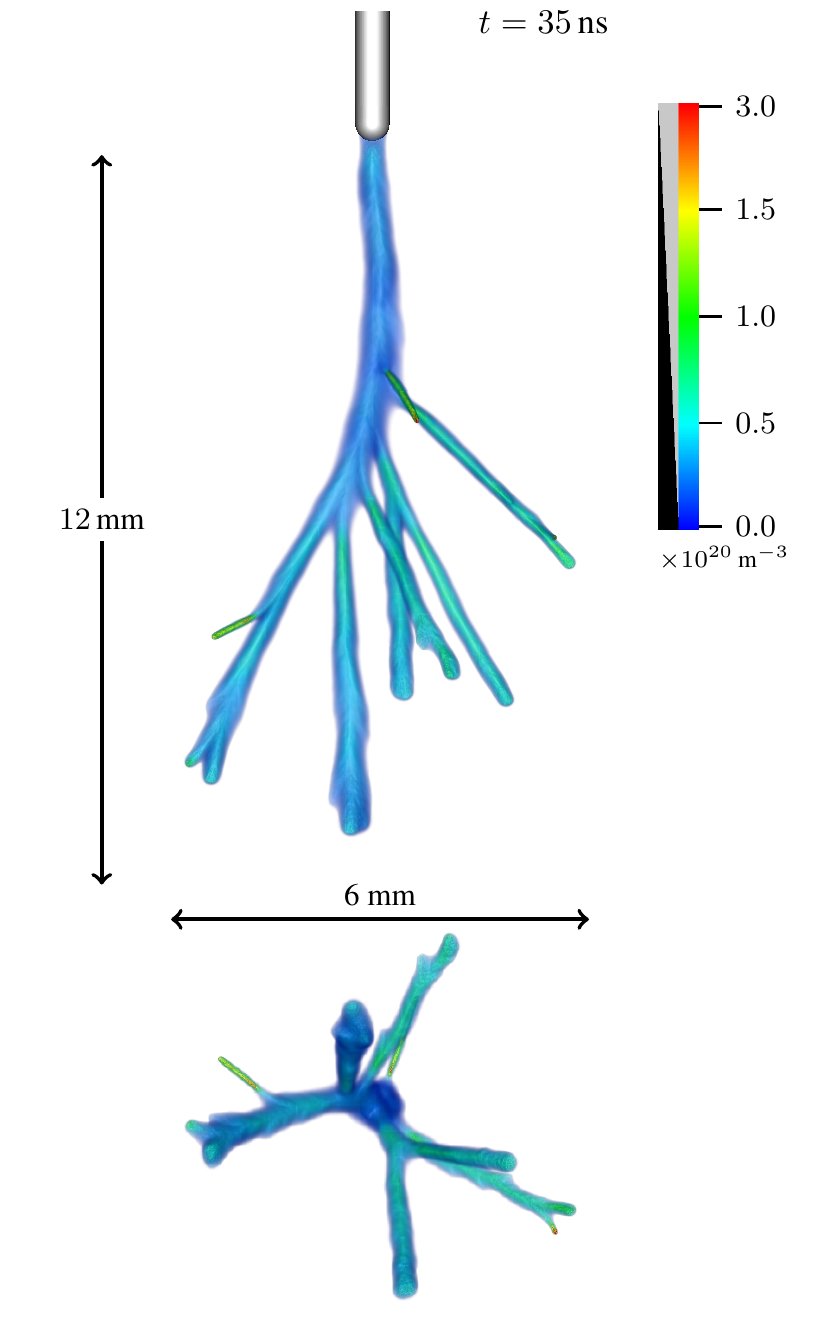}
  \caption{Electron densities computed using the Zheleznyak model \cite{1982TepVT..20..423Z, Pancheshnyi2015}.}
  \label{fig:zheleznyak}
\end{figure}

\section{Results with the Zheleznyak photoionization model}
Next, we provide a computer simulation which used the Zheleznyak photoionization model as discussed in \sref{sec:zheleznyak_model}.
The results are presented in \fref{fig:zheleznyak}.
Recently, a similar study was reported by \citet{Bagheri_2019} and we would like to point out some important differences between that study and the results in this section.
Firstly, we use a field-dependent expression for the excitation efficiency of nitrogen, and this generally leads to generation of fewer ionizing photons.
Secondly, we consider a non-uniform discharge gap with a lower background field, which also leads to generation of fewer photons.
Here, although the average electric field in the discharge gap is $10\,\kV/\cm$, the field is much stronger near the anode and correspondingly also even weaker further into the gap.
Our current simulation model is a fairly realistic representation of the conditions in laboratory discharge experiments. 

\Fref{fig:zheleznyak} shows the simulation results computed with the Zheleznyak model, viewed from two different angles:
From the side, and from below and looking into the streamer.
We observe 10 individual streamer branches, most of which have different thicknesses.
Qualitatively, we observe a comparatively similar discharge evolution as with the \citet{Stephens2018} model.
Both models lead to discharge trees, but the smallest observed streamer diameter with the Zheleznyak model is slightly larger at $150\,\um$ and the highest field at the (thinnest) streamer tip is approximately $250\,\kV/\cm$, which is in fair agreement with the simulations performed using the \citet{Stephens2018} model.
We remark that the computer results are stochastic and that it is currently not known if the observed streamers are minimum-width streamers. 

\section{Final remarks and conclusion}
In summary, we have presented a three-dimensional computational study of positive streamer discharges in air under experimentally available conditions.
Two photoionization models are used, a microscopically based model by \citet{Stephens2018} and the classical Zheleznyak model \cite{, 1982TepVT..20..423Z} with corrections by \citet{Pancheshnyi2015}.
Field-dependent excitation efficiencies were used for both models.
The computer results are qualitatively consistent with experiments and demonstrate a multiscale morphology that includes streamer fluctuations and streamer branching.
Our results demonstrate that streamer branching occurs stochastically, in agreement with previous results \cite{Bagheri_2019}.
Less branching was observed when artificially increasing the photoionization level.
Branching also leads to formation of a discharge tree with thin streamers that carry electric fields exceeding $250\,\textrm{kV/cm}$ at their tips.

Both radiative models still contain uncertainties that are presently difficult to quantify.
The model by \citet{Stephens2018} contains uncertainties due to the accumulated effect of neglected photoionization-capable mechanisms, e.g. excited states where data is not available, and emission from atomic nitrogen and oxygen.
Pending an improvement in the quantification of the collisional quenching rates for the $\Ntwo$ singlet states, an approximate quenching rate has been used. 
The Zheleznyak model \cite{1982TepVT..20..423Z} is derived under equilibrium conditions that are not met in streamer discharges, and the model does not describe experiments with sufficient accuracy \cite{Pancheshnyi2015}.
Furthermore, the use of a single collisional quenching pressure ($40\,\textrm{mbar}$ for atmospheric air) for the Zheleznyak model is questionable as it leads to indiscriminate suppression of all photoionization mechanisms with the same factor.
Microscopic evaluations, however, show that collisional quenching do not affect all transitions equally.
By comparing the photoionization rate in the two models, we found that the Zheleznyak model underestimates the number of short-range photons and overestimates the number of long-range photons.
Nonetheless, no statistical comparison between the two models have been conducted in our study, and we therefore cannot assert whether or not the two models lead to statistically different results.

In the past, several authors have investigated the role of photoionization, predominantly through use of photon continuum approximations in combination with fluid models (see e.g. \cite{Bagheri2018,Naidis2018a} and references therein).
However, continuum models lead to artificial smoothing of the electron distribution in front of the streamer and may lead to suppression of stochastic features such as branching.
The simulations presented in this paper leave no doubt about the importance of discrete photons, as well as the need for precise model parameters.

\section*{Acknowledgements} 
This work was financially supported by the Research Council of Norway through project 245422 and industrial partner ABB AS, Norway.
The computations were performed on resources provided by UNINETT Sigma2 - the National Infrastructure for High Performance Computing and Data Storage in Norway.
The author thanks Dr. J. Teunissen for valuable feedback during preparation of this paper.

\bibliography{references}

\end{document}